\documentclass{elsart}

\usepackage{epsfig}

\usepackage{amssymb}

\newcommand{\Pdz}   {\mbox{\ensuremath{\mathrm{D^0}}}}                        
\newcommand{\Pjpsi} {\mbox{\ensuremath{\mathrm{J}/\psi}}}                      
\newcommand{\dzmm}  {\mbox{\ensuremath{\mathrm{D}^0 \rightarrow \mu^+\mu^-}}}     
\newcommand{\jpsimm} {\mbox{\ensuremath{\mathrm{J}/\psi \rightarrow \mu^+\mu^-}}}  

\newcommand{\Sdz}   {\mathrm{D^0}}                        
\newcommand{\Sjpsi} {\mathrm{J}/\psi}                      
\newcommand{\Tdzmm}  {$\mathrm{D}^0 \rightarrow \mu^+\mu^-$}     

\newcommand{\Br} {\mbox{\ensuremath{\mathcal{B}}}}                      

\newcommand{\del}[1] {}

\begin{document}             

\begin{frontmatter}

\hspace{11.3cm} DESY-04-086\\
\hspace{10.8cm} hep-ex/0405059\\



\title{
Search for the Flavor-Changing Neutral Current Decay \Tdzmm\ with the HERA-B Detector
}

\author[MUE]{{\normalsize I.~Abt}},
\author[DOR]{{\normalsize M.~Adams}},
\author[HAD]{{\normalsize H.~Albrecht}},
\author[ZEU]{{\normalsize A.~Aleksandrov}},
\author[COI]{{\normalsize V.~Amaral}},
\author[COI]{{\normalsize A.~Amorim}},
\author[HAD]{{\normalsize S.~J.~Aplin}},
\author[KIE]{{\normalsize V.~Aushev}},
\author[HAD,VTB]{{\normalsize Y.~Bagaturia}},
\author[MOI]{{\normalsize V.~Balagura}},
% ...
\address[COI]{LIP Coimbra, P-3004-516 Coimbra,  Portugal~\thanksref{GCO}}
% ...
\address[DOR]{Institut f\"ur Physik, Universit\"at Dortmund, D-44221 Dortmund, Germany~\thanksref{GBM}}
% ...
\address[HAD]{DESY Hamburg, D-22603 Hamburg, Germany}
% ...
\address[KIE]{Institute for Nuclear Research, Ukrainian Academy of Science, 03680 Kiev, Ukraine~\thanksref{GKI}}
% ...
\address[MOI]{Institute of Theoretical and Experimental Physics, 117259 Moscow, Russia~\thanksref{GMO}}
\address[MUE]{Max-Planck-Institut f\"ur Physik, Werner-Heisenberg-Institut, D-80805 M\"unchen, Germany~\thanksref{GBM}}
% ...
\address[ZEU]{DESY Zeuthen, D-15738 Zeuthen, Germany}
% ...
\address[VTB]{visitor from High Energy Physics Institute, 380086 Tbilisi, Georgia}
\thanks[GAM]{supported by the Foundation for Fundamental Research on Matter (FOM), 3502 GA Utrecht, The Netherlands}  
\thanks[GBA]{supported by the CICYT contract AEN99-0483}    
\thanks[GRF]{supported by the German Research Foundation, Graduate College GRK 271/3}    
\thanks[GBM]{supported by the Bundesministerium f\"ur Bildung und Forschung, FRG, under contract numbers 05-7BU35I, 05-7DO55P, 05-HB1HRA, 05-HB1KHA, 05-HB1PEA, 05-HB1PSA, 05-HB1VHA, 05-HB9HRA, 05-7HD15I, 05-7MP25I, 05-7SI75I } 
\thanks[GUS]{supported by the U.S. Department of Energy (DOE)}  
\thanks[GCO]{supported by the Portuguese Funda\c c\~ao para a Ci\^encia e Tecnologia under the program POCTI} 
\thanks[GCP]{supported by the Danish Natural Science Research Council}  
\thanks[GKI]{supported by the National Academy of Science and the Ministry of Education and Science of Ukraine} 
\thanks[GLO]{supported by the U.S. National Science Foundation Grant PHY-9986703}  
\thanks[GMO]{supported by the Russian Ministry of Education and Science, grant SS-1722.2003.2, and the BMBF via the Max Planck Research Award}  
\thanks[GOS]{supported by the Norwegian Research Council} 
\thanks[GZU]{supported by the Swiss National Science Foundation}

\begin{abstract}
  We report on a search for the flavor-changing neutral current decay
  \dzmm\ using $50\,\times\,10^6$ events recorded with a dimuon trigger
  in interactions of 920\,GeV protons with nuclei by the HERA-B
  experiment.  We find no evidence for such decays and set a 90$\%$
  confidence level upper limit on the branching fraction
  $\Br(\dzmm)\,<2.0\,\times\,10^{-6}$.
\end{abstract}

\begin{keyword}
Decays of charmed mesons \sep FCNC
  \PACS 13.20.Fc \sep 14.40.Lb 
\end{keyword}

\end{frontmatter}


\section {Introduction}
The decay \dzmm\ \footnote{In this paper, the symbol \Pdz\ denotes
  both \Pdz\ and $\overline \mathrm{D^0}$ mesons.}  is sensitive to
flavor-changing neutral currents (FCNC), which, due to the GIM
mechanism~\cite{GIM}, are forbidden at lowest order and strongly
suppressed at second order in the Standard Model (SM).  In the SM, the
expected contribution of short distance processes to the branching
fraction for this decay is of the order of
$10^{-19}$~\cite{Gorn,Pakvasa}, well below the sensitivity of current
experiments. Long distance effects may enhance the branching fraction
to roughly $10^{-13}$~\cite{Pakvasa,Burdman}, still undetectable by
foreseeable experiments.  However, several extensions of the SM
including the Minimal Supersymmetric Standard Model (MSSM), models
with multiple Higgs doublets, with horizontal gauge bosons, and with
extra fermions, predict an enhancement of the branching fraction by
several orders of magnitude. A comprehensive study of this issue has
been presented in Burdman~et~al.~\cite{Burdman} (see also references
therein). According to this report, an MSSM variant with R-parity
violation predicts the highest estimated branching fraction,
$3.5\,\times\,10^{-6}$.  Thus, the value of the branching fraction of
the \dzmm\ decay mode is exquisitely sensitive to physics beyond the
SM.  Recently, the CDF Collaboration has published a new upper limit
of $2.5\times10^{-6}$ at the 90\% confidence level~\cite{CDF}.

In this letter we report on a search for the \dzmm\ decay using
$50\,\times\,10^6$ events recorded with a dimuon trigger in
interactions of 920\,GeV protons with nuclei in the experiment
HERA-B~\cite{HERAB}. The data were recorded during the 2002-2003 HERA
running period. The branching fraction computation relies on normalizing
the number of events in the \Pdz\ signal region to the number of
reconstructed \jpsimm\ events~\cite{Jing} since possible biases
arising from the dimuon trigger and muon identification largely
cancel.  Thus, for proton interactions on a single target of atomic
weight $A$, the branching fraction limit can be determined using the
following inequality:

\begin{equation}
\Br(\dzmm) \le
\left[\frac{n_{cl}}{N_{\Sjpsi}}\right]
\left[\frac{a_{\Sjpsi}}{a_{\Sdz} \epsilon_{\Sdz}}\right]
\left[\frac{\sigma^{pA}_{\Sjpsi}}{\sigma^{pA}_{\Sdz}}\right]
\Br(\jpsimm)
\label{equation}
\end{equation}
where
\begin{itemize}
\item $n_{cl}$ is the upper limit on the number of \dzmm\ decays;
\item
$N_{\Sjpsi}$ is the number of observed \jpsimm\ events;
\item $a_{\Sdz}$ and $a_{\Sjpsi}$ are the acceptances for \dzmm\ and
  \jpsimm after event quality and particle identification cuts are
  applied (i.e., cuts applied to both channels);
  
\item $\epsilon_{\Sdz}$ is the efficiency for \dzmm\ of cuts designed
  to select secondary vertices (i.e., cuts not applied to \jpsimm);

\item $\sigma^{pA}_{\Sdz}$ and $\sigma^{pA}_{\Sjpsi}$
   are the production cross sections per
  target nucleus for \Pdz\ and \Pjpsi;
\item $\Br(\jpsimm) = (5.88\,\pm\,0.10)\%$~\cite{PDG} is the branching fraction for \jpsimm.

\end{itemize}

The terms $a_{\Sdz}$, $a_{\Sjpsi}$ and $\epsilon_{\Sdz}$ are evaluated
with a complete Monte Carlo simulation. Equation~(\ref{equation}) also requires
knowledge of the relative production cross sections of \Pdz\ and
\Pjpsi . Large errors on the available measurements of the \Pdz\ cross
section dominate the systematic error.

\section{The Detector and Trigger}

\begin{figure}[b]
\epsfig{file=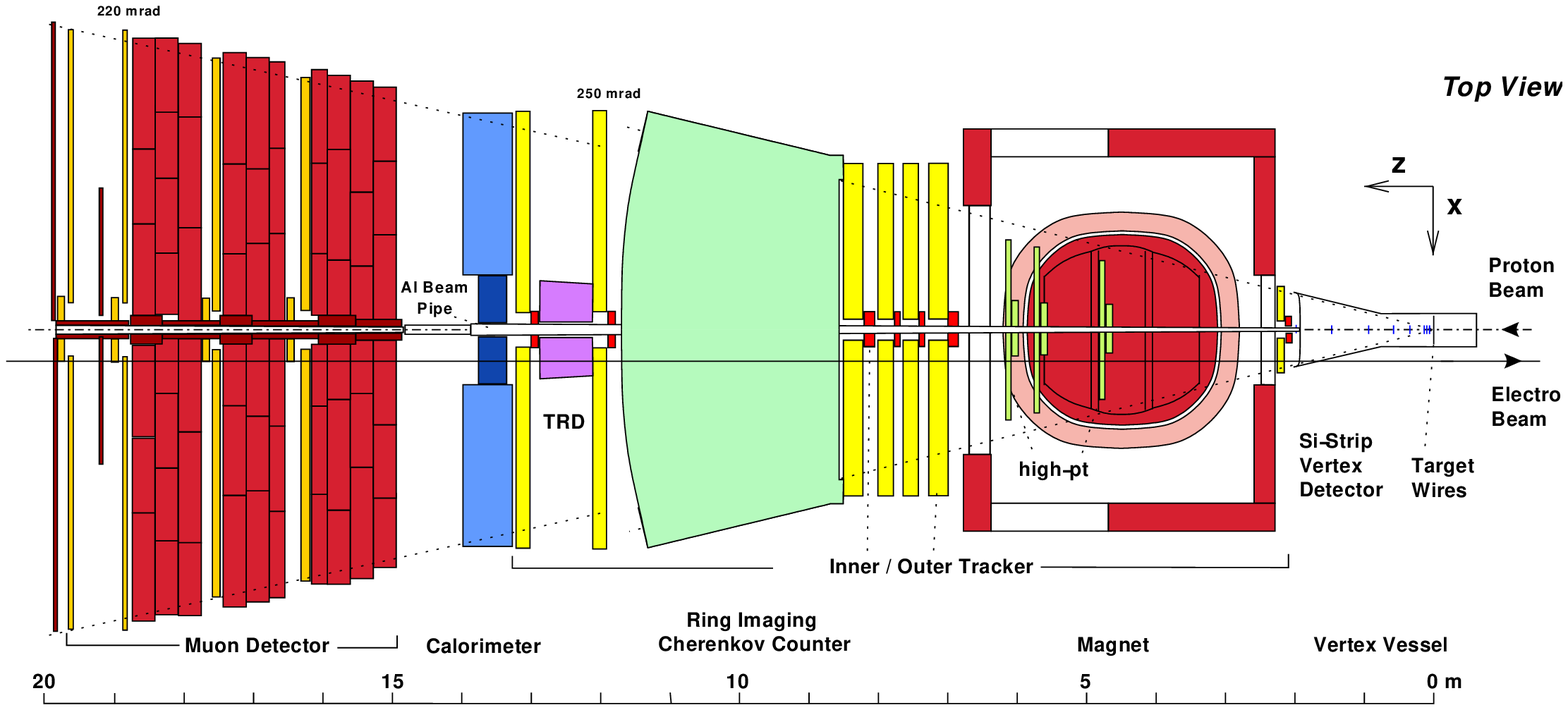,width=.98\columnwidth}
 \caption{\small Plan view of the HERA-B detector.
    \label{fig:layout} }
\end{figure}

The HERA-B fixed-target spectrometer~\cite{HERAB} operates at the
920\,GeV proton beam of the HERA storage ring at DESY and features a vertex
detector and extensive tracking and particle identification systems. It
has a large geometrical coverage from 15 mrad to 220 mrad in the
bending plane and 15 mrad to 160 mrad in the vertical plane. Figure~\ref{fig:layout}
shows a plan view of the detector in the configuration of the 2002-2003
data run.

The target system~\cite{target} consists of two stations of four wires
each. The wires are positioned above, below, and on either side of the
beam and are made from various materials including carbon, titanium,
and tungsten. The stations are separated by 40~mm along the beam
direction.  The wires can be individually moved into the halo of
the HERA proton beam and the interaction rate for each inserted wire
can be adjusted independently.

The Vertex Detector System (VDS)~\cite{vds} is a forward microvertex
detector integrated into the HERA proton ring. It provides a precise
measurement of primary and secondary vertices. The VDS consists of 7
stations (4 stereo views) of double-sided silicon strip detectors
($50\,\times\,70$\,mm, 50 $\mu$m pitch) integrated into a Roman pot system inside a
vacuum vessel and operated as near as 10\,mm from the beam. An additional
station is mounted immediately downstream of the 3\,mm thick aluminum window
of the vacuum vessel.

The first station of the main tracker is placed upstream of the
2.13\,T-m spectrometer dipole magnet. The remaining 6 tracking
stations extend from the downstream end of the magnet to the
electromagnetic calorimeter (ECAL) at 13\,m downstream of the target.
Each tracking station is divided into inner and outer detectors.  The
large area Outer Tracker (OTR)~\cite{otr} consists of 95,000 channels
of honey-comb drift cells and covers the region starting from 200\,mm
from the beam center. The region starting from the beam pipe and extending
up to the start of the OTR acceptance is covered by
micro-strip gas chambers with GEM foils
(Inner Tracker or ITR~\cite{ITR}).

Particle identification is performed by a Ring Imaging Cherenkov
detector (RICH)~\cite{rich}, the
ECAL~\cite{ecal} and a muon detector (MUON)~\cite{muon}.  The MUON
detector is segmented into four superlayers. Iron and concrete shielding extends from
just behind the ECAL to the penultimate MUON superlayer, except for
gaps for the superlayers themselves.
The first two superlayers consist of three layers of tube chambers with
different stereo angles. The last two superlayers each consist of one layer
of pad chambers.

For the sample considered here, triggers are initiated by coincidences
of pad chamber hits (``pretriggers'') from the third and last MUON
superlayers ~\cite{mu-pret}.  Starting from the pretrigger
coordinates, the First Level Trigger (FLT)~\cite{daq} searches for
tracks of charged particles in the MUON tracking layers and four of
the main tracker stations (OTR only).  The momentum of a candidate
track is inferred by assuming the track originates at the target wires
and calculating the bending angle within the magnetic field.
Triggered events are required to have at least one FLT track and at
least one additional pretrigger candidate.

The Second Level Trigger (SLT)~\cite{slt} processor farm receives the
track and pretrigger candidates from the FLT. Starting from regions of
interest defined by the pretriggers, it searches for tracks in the
main tracker, extrapolates those found through the magnet and
attempts to follow them through the VDS. Events with at least two
fully reconstructed tracks which form a good vertex are triggered.
Events passing the SLT are assembled and transferred to a fourth level
farm (no third level algorithm is applied), where a fraction of the
events is fully reconstructed online to provide monitoring.
No event selection is applied at the fourth level.
The final archiving rate is about 100\,Hz.

\section{Monte Carlo Simulation}
\label{MC}

A Monte Carlo (MC) simulation is used to determine the \dzmm\ and
\jpsimm\ acceptance and efficiency terms in equation~(\ref{equation})
and to estimate the background due to $\mathrm{D^0\rightarrow K^-
  \pi^+}$ and $\mathrm{D^0\rightarrow\pi^+\pi^-}$ decays.  The Monte
Carlo events for $pA \rightarrow \Pdz + X$ are generated in two
steps. First, a $c\overline{c}$ pair is generated with PYTHIA 5.7 and
hadronized with JETSET 7.4~\cite{PYTHIA} such that a \Pdz\ is always
produced.  The fractional longitudinal momentum ($x_F$) distribution
of the \Pdz\ is forced into the form $(1-|x_F|)^n$ and its transverse
momentum ($p_T$) distribution is forced into the form $\exp(-b p_T^2)$
by an appropriate selection of events.  For $n$ and $b$, we use
averages of measurements by E653~\cite{E653} and E743~\cite{Ammar}: $n
= 7.7\,\pm\,1.4$, $b = 0.83\,\pm\,0.07\,(\mathrm{GeV}/c)^{-2}$.  After
the generation of the \Pdz\ is completed, the remaining energy is
given as input to FRITIOF 7.02~\cite{FRITIOF}, which generates the
underlying event taking into account further interactions inside the
nucleus. The generated event conserves momentum.

A similar two-step procedure is used to generate the \jpsimm\ sample,
differing only in the method for achieving agreement of \Pjpsi\
kinematic distributions with previous measurements.  In this case,
the generated events are re-weighted according to the
parameterizations of the \Pjpsi\ differential cross sections in
proton-gold collisions at 800\,GeV by E789~\cite{Schub}: $d
\sigma/ d{p_T^2}~\propto~\left(1 + \left(p_{T}/B\right)^2\right)^{-6}$
and $d \sigma/ d{x_F}~\propto~(1 - |x_F| )^n$, with $B= 3.00\pm
0.02\,\mathrm{GeV}/c$ and $n= 4.91\pm 0.18$.

The detector response is simulated with the GEANT 3.21
package~\cite{GEANT}.  Realistic detector efficiencies, readout noise
and dead channels are taken into account. The simulated events are
processed by the same trigger and reconstruction codes as the data.
We have checked that the mis-identification probabilities of pions
from $\mathrm{K}^0_S$ decay, kaons from $\phi(1020)$ decay and
protons from $\Lambda$ decay estimated by the Monte Carlo agree with
those measured in the data.

\section{The Data Sample and Analysis}
\label{sec:anal}

\subsection{Introduction}
\label{sec:anal:intro}
During data-taking, the interaction rate was maintained at
approximately 5\,MHz, resulting in an average of
approximately 0.6~interactions per filled HERA bunch.  A total of
$50\,\times\,10^{6}$ triggers were recorded, with 38\% from runs with
one or two carbon targets, 55\% from runs with carbon and tungsten
targets operated simultaneously, 5\% from runs with one titanium wire
and 2\% from runs with a single tungsten wire.  After eliminating runs
with problematic detector performance, poor beam conditions or
non-standard trigger conditions, $47\,\times\,10^{6}$ events remain.

The events are reconstructed with the standard HERA-B analysis package
~\cite{cats,track}. The search for \Pjpsi\ and \Pdz\ candidates is
performed by analyzing the mass spectrum for unlike-sign dimuon
candidates. \Pdz\ candidates are further selected by requiring a
detached vertex. The number of background events in the signal region
is calculated from sidebands. The upper limit is obtained by comparing
the observed number of events inside the signal region with the
background estimate and normalizing to the number of observed
\Pjpsi's, after appropriate corrections are applied.

\begin{figure}[b]
\setlength{\unitlength}{1mm}
\begin{center}
\epsfig{file=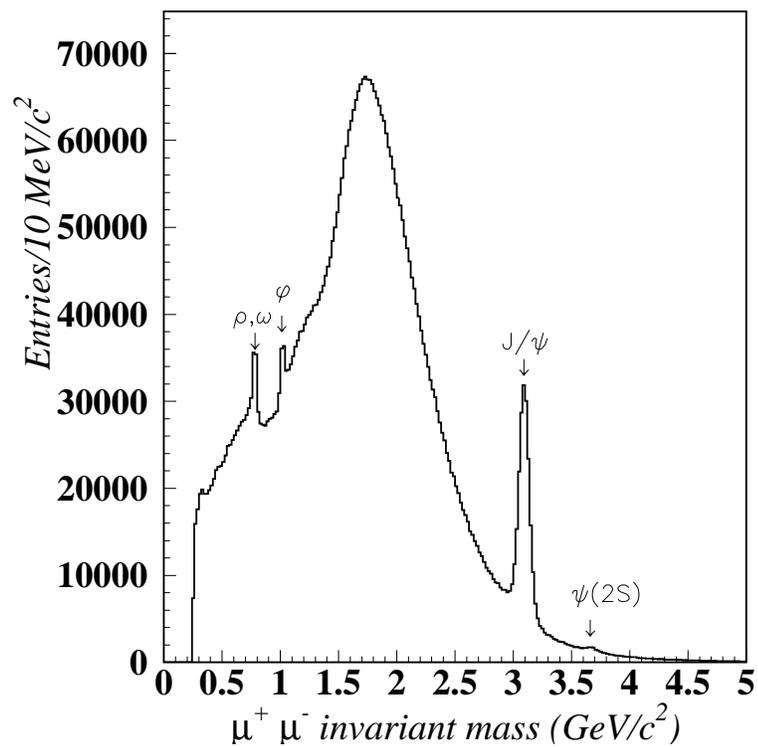, width=0.7\columnwidth, bbllx=30,bblly=130,bburx=530,bbury=660}
 \caption{\small Invariant mass of unlike-sign dimuon candidates after minimal cuts. The
   positions of known resonances are indicated.
    \label{fig_01} }
\end{center}
\end{figure}

The mass spectrum for unlike-sign dimuon candidates after minimal cuts
is shown in Fig.~\ref{fig_01}. Peaks at the $\omega/\rho$, $\phi$,
\Pjpsi\ and $\psi (2S)$ masses are clearly visible.  The fall-off
towards low mass is caused by an implicit $p_T$ cut of about
$0.7\,\mathrm{GeV}/c$ imposed by the SLT when defining regions of
interest for subsequent tracking.  A fit in the mass interval
2.4--3.5$\,\mathrm{GeV/c^2}$ to an exponential background plus a
Gaussian with a radiative tail yields $147710\pm520$ \Pjpsi\ decays in
a two standard deviation window centered on the mean value from the
Gaussian fit: $3.095\,\mathrm{GeV}/c^2$. The fitted width is
$44\,\mathrm{MeV}/c^2$. The mean value is 1.9\,$\mathrm{MeV}/c^2$
below the PDG value~\cite{PDG}.  From these numbers we estimate the
mass resolution in the \Pdz\ region to be about 25\,$\mathrm{MeV}/c^2$
and that the reconstructed mass of a \dzmm\ signal would be within
$1.1\,\mathrm{MeV}/c^2$ of the PDG~\cite{PDG} value.

\subsection{Event Selection}
\label{sub:evs}

The main background for the \dzmm\ channel results from muon pairs
from independent $\pi^\pm$ or $K^\pm$ decays which appear to form a
secondary vertex displaced from the primary vertex.  The cuts are designed to
minimize this background while maintaining high efficiency for \Pdz\ decays.

The data sample is first reduced by application of relatively loose
cuts. In this first pass, a general vertex-finding algorithm is
applied and events with at least one reconstructed primary vertex and
at most one reconstructed primary vertex per target wire are selected.
Muon candidate tracks are then selected by requiring that the muon
probability, $\mathcal{P}_\mu$, derived from the MUON hit information,
be greater than 0.01, that the $\chi^2$ per degree of freedom of the
track fit ($\chi^2_{tr}/d.o.f.$) be less than 20, and that the kaon
identification likelihood probability from the RICH reconstruction be
less than 0.4.  The two muon tracks are then excluded from the primary
vertex, and primary and dimuon vertices are fitted.  The
$\chi^2$~probability of the primary vertex is required to be greater
than $0.01$ and that of the dimuon vertex greater than $0.2$.  The
number of tracks per primary vertex is required to be less than 50.
Approximately 0.93~million dimuons with mass in the \Pdz\ region and
69000 \Pjpsi\ decays (estimated here and below using the fit described
in Sec.~\ref{sec:anal:intro}) survive the initial cuts.

In the following, we call the muon pair with a fitted secondary vertex
a ``dimuon pseudoparticle''.  We consider two regions
of $\mu^+\mu^-$ invariant mass: $2.7 - 4.0\,\mathrm{GeV}/c^2$, which
we refer to as the ``\Pjpsi\ region'', and $1.59 -
2.15\,\mathrm{GeV}/c^2$, which we refer to as the ``\Pdz\
region''.

Further selection cuts are divided into two parts:
\begin{itemize}
\item Common cuts applied both for \Pdz\ and \Pjpsi\ regions.  These
  cuts are mainly quality cuts, the efficiencies of which are nearly
  identical for muons from possible \dzmm\  and for muons from
  \jpsimm. Small differences in the efficiencies coming from the
  different momentum distributions of muons from \Pdz\ decay and
  \Pjpsi\ decay are evaluated from the data and Monte Carlo.
 \item Cuts applied only for the \Pdz\ region (lifetime cuts).  These
  cuts are intended to select a well-defined detached secondary vertex
  associated with the selected primary vertex.
\end{itemize}

The following common selection criteria are applied for \Pdz\ and
for \Pjpsi\ regions:
\begin{itemize}
\item a cut on total track multiplicity ($N_{tr}$) to suppress
multi-event pile-up which is enhanced by the two-muon requirement;
\item a cut on $\chi^2_{tr}/d.o.f.$ for each
  muon to suppress ghosts and $\pi$/$\mathrm{K}$ decays in flight;
\item a cut on $\mathcal{P}_\mu$ to reduce fake dimuon events;
\item a cut on the transverse momentum of each muon ($p_T^\mu$) to
  suppress muons from pion and kaon decays.
\end{itemize}

To optimize the common cuts, we employ a blind analysis technique: all
dimuons from the \Pdz\ signal region are masked and the cuts are chosen to
maximize the quantity $N_{\Sjpsi}/\sqrt{B_{\Sdz}}$, where $N_{\Sjpsi}$
is the number of \Pjpsi\ candidates above background found in the
$\mu^+\mu^-$ invariant mass spectrum in a two standard deviation
window around the \Pjpsi\ position and $B_{\Sdz}$ is the expected
background in the \Pdz\ signal region ($1.815-1.915~\,\mathrm{GeV}/c^2$),
estimated from \Pdz\ sidebands ($1.59~-~1.79\,\mathrm{GeV}/c^2$ and
$1.94~-~2.14\,\mathrm{GeV}/c^2$).  The resulting cuts are:
$N_{tr} < 45$, $\chi^2_{tr}/d.o.f. < 7.5$, $\mathcal{P}_\mu > 0.7$, 
$p_T^\mu > 0.7\,\mathrm{GeV}/c$.

\begin{figure}[t]
\setlength{\unitlength}{1mm}
 \begin{picture}(10,85)
 \put(-8,-15){\includegraphics{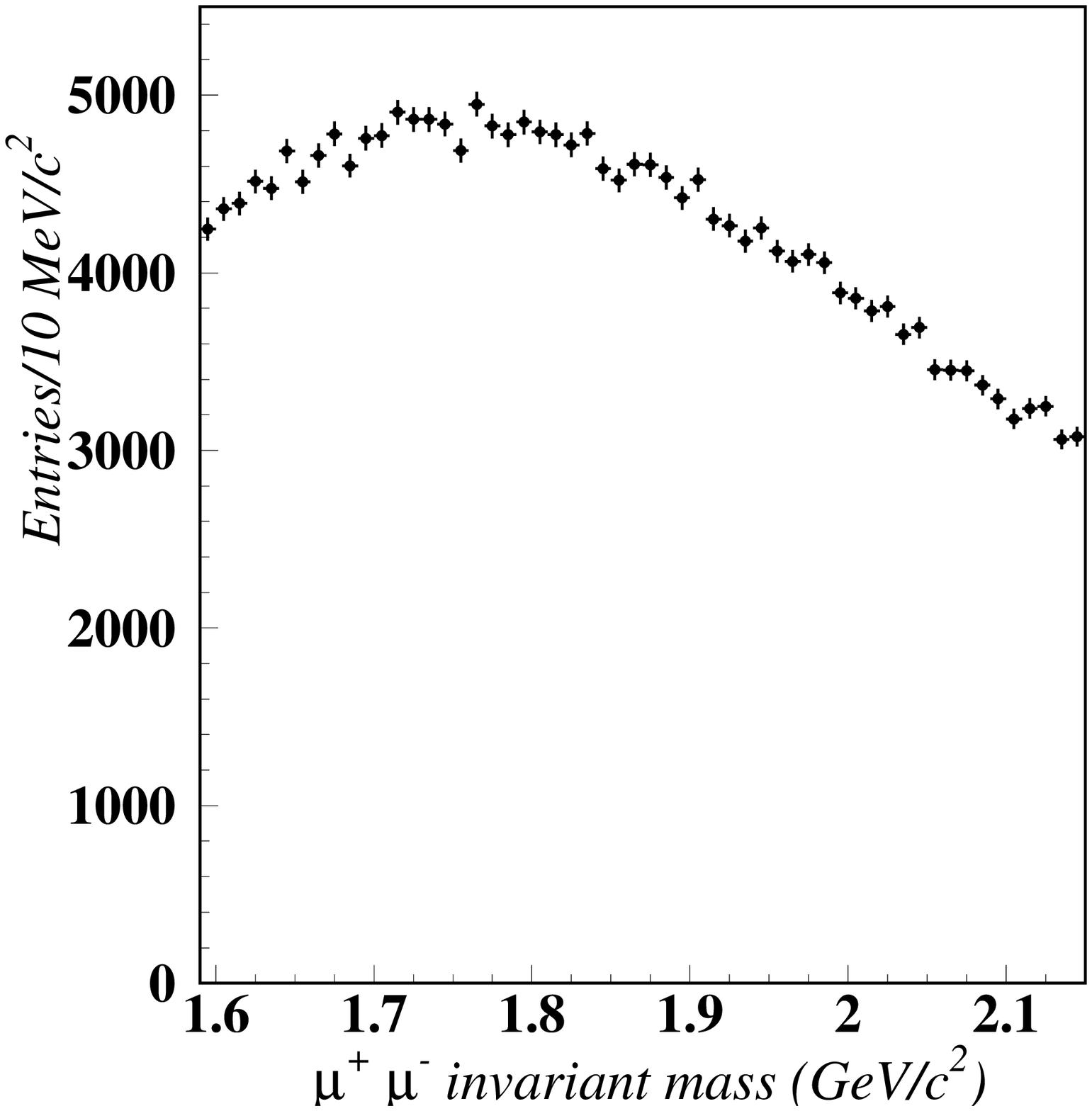}}
 \put(66,-15){\includegraphics{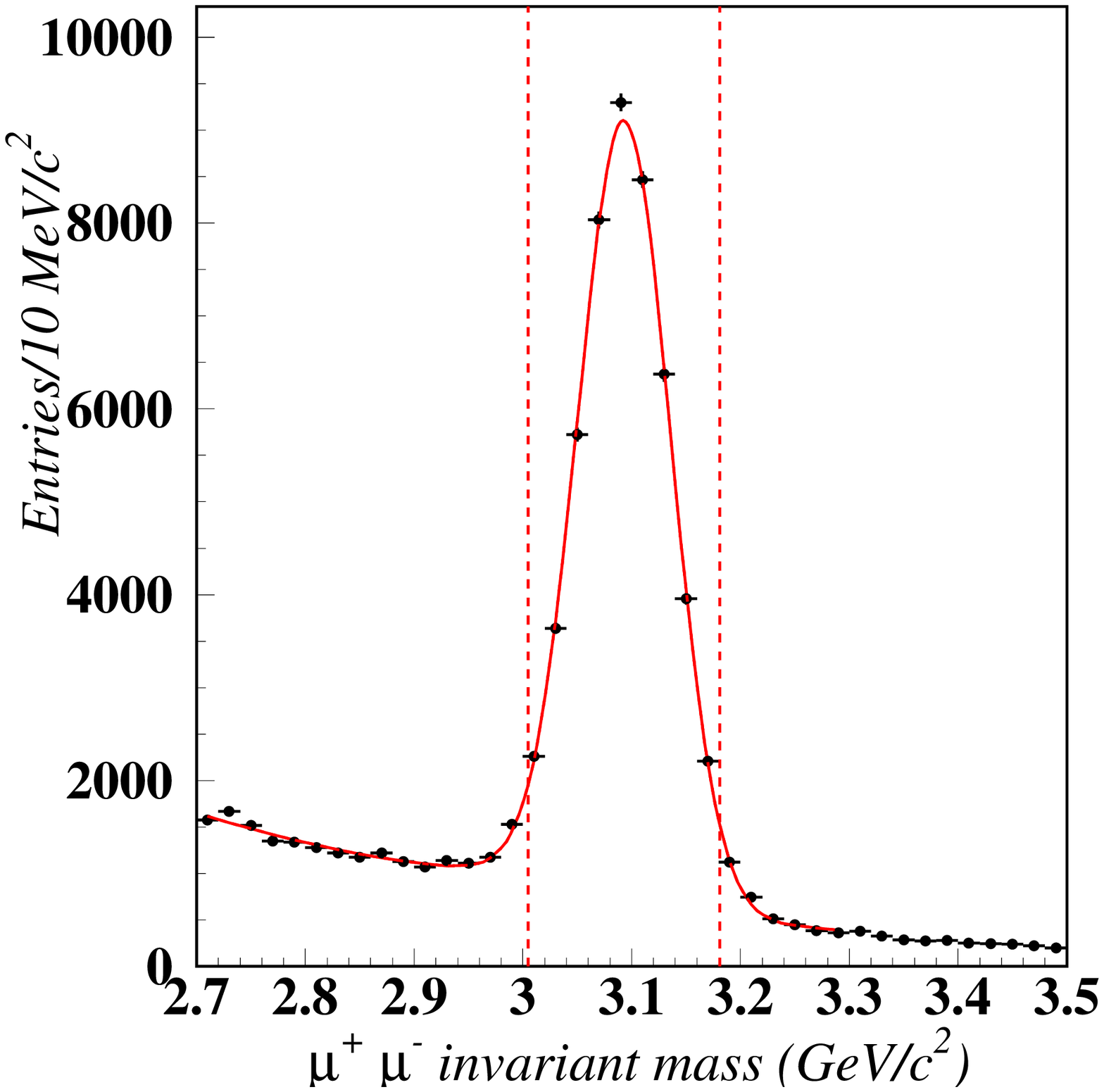}}
 \put(60,70){a)}
 \put(130,70){b)}
 \end{picture}
 \caption{\small Candidates with two oppositely charged muon tracks
   for (a) \Pdz\ and (b) \Pjpsi\ regions after all common (see text) cuts
   have been applied. The
   curve in (b) is the result of a fit to an exponential to describe the
   background and a Gaussian with a radiative tail to describe the
   signal. The two standard deviation selection window around the \Pjpsi\
   position is indicated by the dashed lines.
   \label{fig_02} }
\end{figure}

After all common cuts are applied, about 238000 events in the \Pdz\ region
and 46000 events in the \Pjpsi\ peak remain.  The dimuon mass
distributions for \Pdz\ and \Pjpsi\ regions with the above cuts are
shown in Fig.~\ref{fig_02}.

\subsection{\Pdz\ Selection}

To isolate possible \Pdz\ mesons, cuts are applied to three
quantities: the separation between primary and secondary vertices, the
proper decay length, and the impact parameter of the dimuon
pseudoparticles to the primary vertex.  At HERA-B energies, nearly all
\Pdz\ mesons originate at the target wires, i.e., the fraction arising
from B decays is negligible ($< 0.1\%$).
                
Most \Pdz\ mesons decay within a few millimeters of the primary vertex
in the laboratory frame. This distance is comparable to the precision
of the secondary vertex measurement in the longitudinal direction.  To
ensure that the secondary vertex is well separated from the associated
primary vertex, we compare this distance with uncertainties of primary
and secondary vertices and cut on the separation value $\Delta z =
(z_{sec}-z_{pr})/\sqrt{\sigma^2_{z_{sec}}+\sigma^2_{z_{pr}}}$, where
$z_{pr}$ and $z_{sec}$ are the $z$-coordinate (along the beam direction)
of primary and secondary vertices, respectively, and
$\sigma_{z_{sec}}$, $\sigma_{z_{pr}}$ are their calculated errors.
The average resolutions are $\sigma_{z_{sec}}~=~500\,\mu\mathrm{m}$
and $\sigma_{z_{pr}}~=~420\,\mu\mathrm{m}$.

The proper decay length is given by $c\tau~=~mc\cdot L/p$, where $m$
is the dimuon invariant mass, $L$ is the decay length in the
laboratory frame, and $p$ is the reconstructed momentum of the dimuon
pseudoparticle.

The impact parameter ($I_v$) is defined as the distance between the
primary vertex and the point of intersection of the dimuon
pseudoparticle flight direction with the $xy$ plane at the $z$
position of
the primary vertex.  The impact parameter resolution is typically in
the range of 30\,$\mu$m to 60\,$\mu$m.

To optimize the cuts, we apply a blind analysis technique
similar to that used to optimize the common cuts.  Since these three
quantities are correlated, a three-dimensional optimization is
performed wherein the ratio $N^{MC}_{\Sdz}/S$ is maximized.
$N^{MC}_{\Sdz}$ is the number of reconstructed Monte Carlo events in
the \Pdz\ peak. The ``experimental sensitivity'',$S$, is the average
90\% confidence level upper limit on the number of signal events
obtained for an ensemble of experiments with the expected background
estimated from the \Pdz\ sidebands, assuming no signal from \dzmm\ is
present (from Table~XII of Ref.~\cite{Feldman}).

This optimization procedure produces the following cuts: $\Delta
z > 7.0 $, $c\tau > 250\,\mu$m and $ I_v < 110\,\mu$m.
The optimized experimental sensitivity is $S_{opt} = 5.53$ events.

\section{Results}

\begin{figure}[p]
\setlength{\unitlength}{1mm}
 \begin{picture}(10,85)
 \put(22,-15){\includegraphics{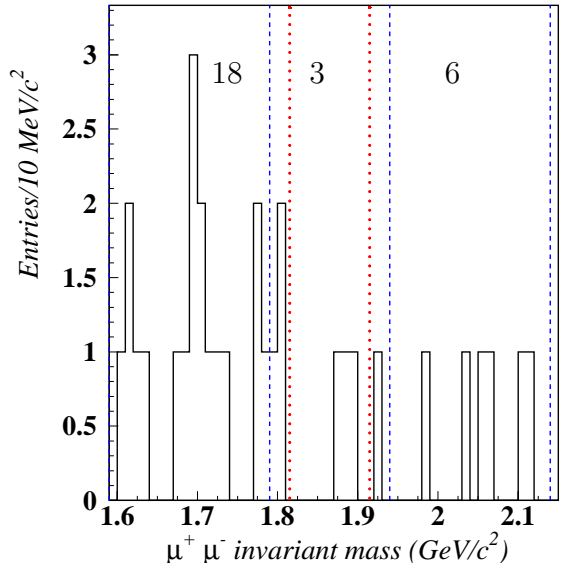}}
 \put(53,70){18}
 \put(66,70){3}
 \put(84,70){6}
\end{picture}
 \caption{\small Unlike-sign dimuon mass spectra in the \Pdz\ region
   after all cuts have been applied. The sidebands are indicated by
   dashed lines and the signal region is indicated by dotted lines. The
   numbers of events in each of the regions are given.
    \label{fig_03} }
\end{figure}

After applying all cuts, 31 events remain in the \Pdz\ mass region as
shown in Fig.~\ref{fig_03}.  The sidebands are indicated by dashed
lines and signal region is indicated by dotted lines. The signal
region contains 3 events.

The number of background events in the \Pdz\ mass region from
$\mathrm{D^0} \rightarrow \mathrm{K}^- \pi^+$ decays is estimated from
Monte Carlo to be $1.8\pm1.0$.  All other \Pdz\ decay modes give a
negligible contribution.  The background shape is therefore not
significantly influenced by charm decay and we estimate the background
using the shape of the mass plot before \Pdz\ selection cuts are
applied (see Fig.~\ref{fig_02}a).  After normalizing to the number of
events in the sidebands, the expected background in the signal region
is $6.0\,\pm 1.2$.  We have checked that the background shape does not
significantly change relative to the uncut distribution when one of
the three selection cuts is removed (three tests). The maximum
difference between background estimates derived from the uncut
distribution and the three partially cut distributions is 4\%. We note
that a simple linear interpolation between sidebands also predicts
$6.0\,\pm\,1.2$ background events in the signal region.

The upper limit calculation for the \dzmm\ decay mode requires
knowledge of the \Pdz\ to \Pjpsi\ production cross section ratio in
proton-nucleus interactions.  Three experiments have published results
on the \Pdz\ production cross section at 800\,GeV.  The $x_F$
coverages of experiments E653~\cite{E653} and E743~\cite{Ammar} are
similar to that of HERA-B: $-0.2\,<\,x_F\,<\,0.1$. The $x_F$ coverage of
experiment E789~\cite{Leitch94} is relatively small and the E653 and
E743 parameterizations have been used to extrapolate to the full $x_F$
range. For this reason we do not include it in the calculation of
the \Pdz\ cross section. From the remaining two
measurements of $22^{+9}_{-7}\pm5\,\mathrm{\mu b/nucleon}$ and $38\pm
3\pm 13\,\mathrm{\mu b/nucleon}$ we obtain
$\sigma^{pN}_{\Sdz} = 27.3\pm7.7\,\mathrm{\mu b/nucleon}$.

The prompt \Pjpsi\ production cross section per nucleon,
$\sigma^{pN}_{\Sjpsi}$, has been measured by two fixed target
experiments~\cite{Schub,Alex} at 800\,GeV.  Both measurements were
performed with nuclear targets and extrapolated to atomic weight ($A$)
of 1, assuming $\sigma^{pA}_{\Sjpsi} =
\sigma^{pN}_{\Sjpsi} \cdot A^{\alpha_{\Sjpsi}}$.
~~~After adjusting these measurements using the most recent
measurement of $\alpha_{\Sjpsi}= 0.955~\pm~0.005$~\cite{Leitch}, and
averaging them we obtain a prompt \Pjpsi\ production cross section of
$\sigma^{pN}_{\Sjpsi}$ =
$333~\pm~6~\pm26\,\mathrm{nb/nucleon}$ at 800\,GeV.
We assume that the ratio of \Pdz\ and \Pjpsi\ cross sections does
not change significantly between 800\,GeV and 920\,GeV.

The ratio of acceptances, not including cuts applied only for
selecting \dzmm, is $a_{\Sdz}/a_{\Sjpsi} = 0.287\,\pm\,0.028$.  The
error is evaluated by studying the efficiencies of the applied cuts
when detector characteristics (channel efficiencies, maskings,
resolutions) are varied within ranges that reflect their
uncertainties. The quadratic sum of contributions from all significant
sources, including triggering, target geometry, muon and kaon
identification and Monte Carlo statistics is 9.8\%.  The efficiency of
the additional secondary vertex cuts applied in the \Pdz\ region is
$\epsilon_{\Sdz} = (6.83\,\pm\,1.08)\,\times\,10^{-2}$.

Data samples from the various targets are combined into a single data
set assuming the production cross sections per nucleon depend on
atomic weight ($A$) as $\sigma^{pA}_{\Sdz(\Sjpsi)} =
\sigma^{pN}_{\Sdz(\Sjpsi)} \cdot
A^{\alpha_{\Sdz\left(\Sjpsi\right)}}$, with
$\alpha_{\Sdz}=1.02\,\pm 0.03\,\pm\,0.02$ (E789)~\cite{Leitch94} and
$\alpha_{\Sjpsi}=0.955\,\pm\,0.005$ (E866)~\cite{Leitch}.

Equation (\ref{equation}) can be rewritten as:

\begin{equation}
\Br(\dzmm) \le \frac{n_{cl}}{F^{sens}}
\label{equation2}
\end{equation}

where
\begin{equation}
F^{sens} \equiv
{\sum_i{\left( N_{\Sjpsi}^i A_i^{\alpha_{\Sdz}- \alpha_{\Sjpsi}}\right) }}
~\left[\frac{a_{\Sdz} }{a_{\Sjpsi}}\epsilon_{\Sdz}\right]
~\left[\frac{\sigma^{pN}_{\Sdz}}{\sigma^{pN}_{\Sjpsi}
~\Br(\jpsimm)}\right]
\label{equation3}
\end{equation}
is the detector sensitivity after summation over targets.

The values and associated errors for all terms in
equation~(\ref{equation3}) are given in Table~\ref{syst}.  Using these
numbers, $F^{sens} = 1.57\,\times\,10^6$.
After combining statistical and systematic
errors quadratically the total relative error on $F^{sens}$ from all
contributing terms in equation~(\ref{equation3}) is 37\%.

\begin{table}[hbt]\centering
\caption{Values and contributions to the systematic uncertainty on
$\Br(\dzmm)$
for the factors in equation~(\ref{equation3}).
  ($^\star$ - statistical and systematic errors added quadratically)}
\hspace{2mm}
\begin{tabular}{|c|c|c|} \hline
factor & value & $\%$ \\ \hline
$a_{\Sdz}/a_{\Sjpsi}$ & $0.287\,\pm\,0.028 $ & 9.8\\
$\epsilon_{\Sdz}$ & $ (6.83 \pm 1.08)\,\times\,10^{-2}$ & 15.8\\
$\sigma^{pN}_{\Sjpsi} (\mathrm{nb/nucleon})$ & $ 333 \pm 6 \pm
26$ & 8.0\\
$\sigma^{pN}_{\Sdz} (\mathrm{\mu b/nucleon})$ & $27.3 \pm
7.7^\star$ & 28.2\\
\Br(\jpsimm) & $ (5.88 \pm 0.10)\,\times\,10^{-2}$ & 1.7\\
$N^{C}_{\Sjpsi}$ & $31010 \pm 200 (stat)$ & 0.7\\
$N^{W}_{\Sjpsi}$ & $12660 \pm 140 (stat)$ & 1.1\\
$N^{Ti}_{\Sjpsi}$ & $2430 \pm 60 (stat)$ & 2.5\\
$\alpha_{\Sdz} - \alpha_{\Sjpsi}$ & $0.065 \pm 0.036^\star$ & 12.3\\
\hline
\end{tabular}
\label{syst}
\end{table}

Recent publications~\cite{stats} employ a variety of methods for
calculating upper limits and there is no universally accepted
procedure~\cite{Feldman,Zech,Narsky}.  We choose an approach similar
to that first advocated by Feldman and Cousins~\cite{Feldman}.  This
method has been since extended by Conrad~et~al.~\cite{Conrad} to
incorporate uncertainties in detector sensitivity and the background
estimate based on an approach described by Cousins and
Highland~\cite{Highland}.  A further refinement of the Conrad~et~al.
method by Hill~\cite{Hill} results in more appropriate behavior of the
upper limit when the observed number of events is less than the
estimated background, as is the case for the present measurement. We
have adopted this method but note that Table~\ref{table_vik} contains
all of the numbers needed to calculate an upper limit using any of the
methods in the papers cited above.  We assume that the probability
density functions of $F^{sens}$ and background estimates are
Gaussian-distributed.

Using equation~(\ref{equation2}) and taking systematic errors and
background fluctuations into account with the Hill
prescription~\cite{Hill2}, the upper limit at 90\% confidence level 
on $\Br(\dzmm)$ is
$2.0\,\times\,10^{-6}$.

\begin{table}[hbt]\centering
\caption{
Summary of parameters entering into the upper limit calculation
}
\hspace{2mm}
\begin{tabular}{|p{70mm}|l|} \hline
Sensitivity factor, $F^{sens}$ & $(1.57\,\pm\,0.58)\times\,10^6$\\
Number of events in the signal region & 3\\
Expected background events & 6.0 $\pm$ 1.2 \\
\hline

\end{tabular}
\label{table_vik}
\end{table}

\section{Conclusions}
\label{Conclusion}

We have investigated the dimuon mass spectrum in a search for
the flavor-changing neutral current decay \dzmm.
Using the values of \Pdz\ and \Pjpsi\ production cross sections
published in the literature we have set an upper limit
on the branching fraction of \\
\begin{center}
$\Br(\dzmm)~<~2.0\,\times\,10^{-6}$
\end{center}
at $90\%$ confidence level.
For comparison, the lowest previously published limit on $\Br(\dzmm)$
is $2.5\,\times\,10^{-6}$ at 90$\%$ confidence level~\cite{CDF}.

According to the recent calculations by Burdman~et~al.~\cite{Burdman},
in the case of strong $R$-parity violation in the MSSM the \dzmm\ 
branching fraction is at the level of $3.5\,\times\,10^{-6}$.  Thus, our
result disfavors this scenario and constrains the size of $R$-parity
violating couplings which enter into the predicted branching fraction.

\section*{Acknowledgments}

We are grateful to the DESY laboratory and to the DESY accelerator
group for their strong support since the conception of the HERA-B
experiment.  The HERA-B experiment would not have been possible
without the enormous effort and commitment of our technical and
administrative staff. We thank R.~D.~Cousins, G.~C.~Hill and G.~Zech
for helpful discussions on procedures for setting confidence limits.



\end{document}